\documentclass[fleqn]{aa54}
\usepackage{graphicx,natbib,txfonts,url}
\bibpunct{(}{)}{;}{a}{}{,} 
\def\cite#1{\citealp{#1}} 
\def\citealias#1{\citetalias{#1}} 

\newcommand{\figureone}[3]{%
\begin{figure}[tbp]
\begin{center}
\includegraphics[width=88mm]{#1}
\caption{#3}
\label{#2}
\end{center}
\end{figure}
}

\newcommand{\figuretwo}[3]{%
\begin{figure*}[tbp]
\begin{center}
\includegraphics[width=180mm]{#1}
\caption{#3}
\label{#2}
\end{center}
\end{figure*}
}

\newcommand{\figurethree}[3]{%
\begin{figure*}[tbp]
\sidecaption
\includegraphics[width=120mm]{#1}
\caption{#3}
\label{#2}
\end{figure*}
}

\def\kf{($k_\mathrm{h},f$)}
\def\CaII{\mbox{Ca\,\sc{ii}}} 
\def\CaIIH{\mbox{Ca\,\sc{ii}\,\,H}}
\def\CIV{\mbox{C\,\sc{iv}}} 
\def\FeI{\mbox{Fe\,\sc{i}}} 
\def\HtwoV{\mbox{H$_{2V}$}}
\def\VminI{\hbox{$V\!\!-\!\!I$}}           
\def\IminI{\hbox{$I\!\!-\!\!I$}}           
\def\VminV{\hbox{$V\!\!-\!\!V$}}           

\defcitealias{2001A&A...379.1052K}{Paper~I} 

\begin{document}

\title{Dynamics of the solar chromosphere} 
\subtitle{V. High-frequency modulation in ultraviolet image sequences 
             from TRACE}
\titlerunning{High-frequency modulation in TRACE images}

\author{A.G.~de~Wijn\inst{1}
        \and
        R.J.~Rutten\inst{1,2}
        \and
        T.D. Tarbell\inst{3}}

\institute{Sterrekundig Instituut, 
           Utrecht University, 
           Postbus~80\,000, 3508~TA~Utrecht, The~Netherlands\\
           \email{A.G.deWijn@astro.uu.nl, R.J.Rutten@astro.uu.nl}
         \and
           Institute of Theoretical Astrophysics, 
           Oslo University, 
           P.O.~Box~1029 Blindern, N-0315 Oslo, Norway
         \and
           Lockheed Martin Solar and Astrophysics Lab,
           Dept.\ ADBS, Building 252, 3251 Hanover Street, 
           Palo Alto, CA 94304, USA\\
           \email{tarbell@lmsal.com}
           }

\date{Received 24 July 2004 / Accepted 12 October 2004}
\offprints{A.G.~de~Wijn,\\
           e-mail: {\tt A.G.deWijn@astro.uu.nl}}

\abstract{We search for signatures of high-frequency oscillations in the
upper solar photosphere and low chromosphere in the context of acoustic
heating of outer stellar atmospheres.  We use ultraviolet image sequences
of a quiet center-disk area from the \emph{Transition Region and Coronal
Explorer} (TRACE) mission which were taken with strict cadence regularity.
The latter permits more reliable high-frequency diagnosis than in earlier
work.  Spatial Fourier power maps, spatially averaged coherence and
phase-difference spectra, and spatio-temporal \kf\ decompositions all
contain high-frequency features that at first sight seem of considerable
intrinsic interest but actually are more likely to represent artifacts of
different nature.  Spatially averaged phase difference measurement provides
the most sensitive diagnostic and indicates the presence of acoustic
modulation up to $f\approx20~\mathrm{mHz}$ (periods down to 50~seconds) in
internetwork areas.  \keywords{Sun: photosphere -- Sun: chromosphere --
Sun: oscillations}}

\maketitle

\section{Introduction}\label{sec:introduction}
In this paper we continue studies of solar atmosphere oscillations based on
analyzing the temporal brightness modulation in image sequences taken with
the \emph{Transition Region and Coronal Explorer} (TRACE) in ultraviolet
passbands which sample the upper solar photosphere and low solar
chromosphere.  We again exploit the absence of seeing in TRACE data (apart
from space-weather particle hits) to provide extensive Fourier diagnostics
for quiet-sun network and internetwork areas with excellent sampling
statistics.

In
	\citet[%
	henceforth \citealias{2001A&A...379.1052K}] 
	{2001A&A...379.1052K}, 
these techniques were used in a comprehensive overview of quiet-sun
brightness oscillation properties derived from TRACE image sequences in its
three ultraviolet passbands centered at $\lambda=1700$, $1600$, and
$1550~\mbox{\AA}$.  In standard models of the solar atmosphere such as FALC of
  \citet{1993ApJ...406..319F} 
these passbands sample layers just below, at, and just above the
temperature minimum, respectively.  The subsequent paper by
	\citet{2003A&A...407..735R} 
compared low-frequency ultraviolet brightness modulation at
these wavelengths to the underlying white-light patterns in quiet-sun
areas.
	\citet{2003A&A...401..685M} 
analyzed TRACE ultraviolet brightness modulation maps containing an active
region.

In this paper we return to the high-frequency aspects of ultraviolet
brightness modulation.  The data used in
	\citealias{2001A&A...379.1052K} 
suffered from irregular timing intervals between successive images,
severely reducing the high-frequency information content.  The sequences
used here have strict sampling regularity and are therefore better suited
to search for high-frequency oscillation signatures.  We also employ a much
improved alignment method.

The obvious motivation for such searches is given by the long quest for
acoustic heating of outer stellar atmospheres started by
	\citet{1948ZA.....25..161B} 
and
	\citet{1948ApJ...107....1S}. 
It is concisely summarized by 
	\citet{2002A&A...395L..51W}, 
to whom we refer for further background.
	\citet{2002A&A...395L..51W} 
employed image sequences from the G\"ottingen Fabry-Perot
spectrometer at the German Vacuum Tower Telescope on Tenerife,
scanning the non-magnetic \FeI\ 5434~\AA\ line which samples layers
around $h=500~\mathrm{km}$ above the white-light surface.
They inferred the presence of sufficient power with
50\,--\,100~s periodicity (10\,--\,20~mHz in frequency) to compensate
the radiative losses of the chromosphere, with apparent spatial power
concentration above intergranular lanes.  In this analysis we
use TRACE data to search for corroborative evidence in
ultraviolet brightness modulation from the same layers.

\section{Observations and data reduction}\label{sec:observations}

\figureone{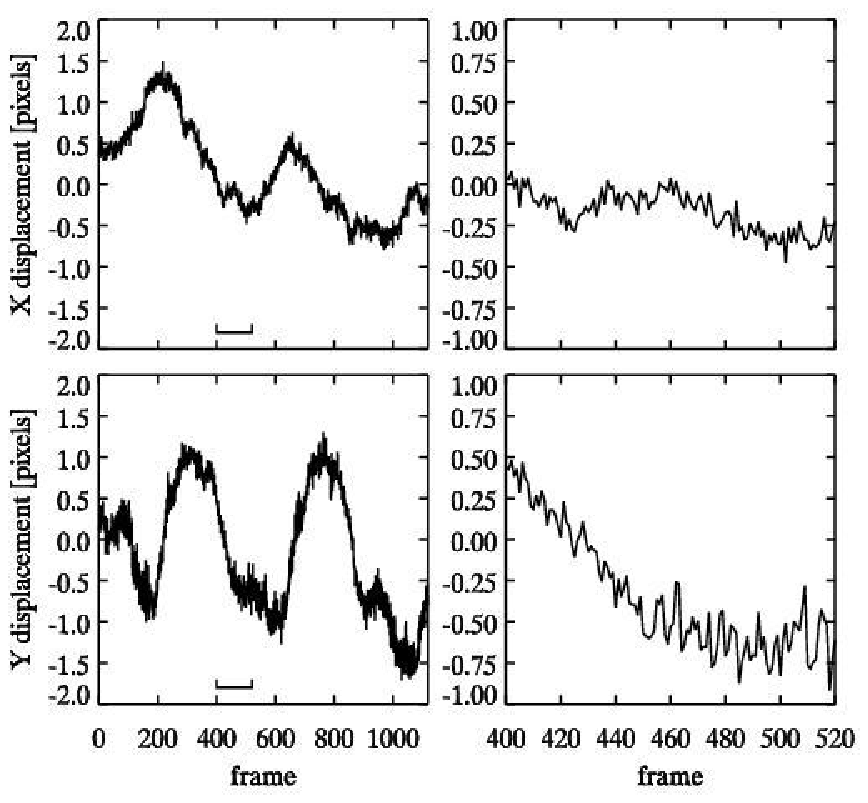}{fig:disp1600}{Corrections for residual image
displacements for the 1600-\AA\ sequence, in the solar $X$ (upper panels)
and $Y$ directions (lower panels)} plotted against frame number.  The
enlargements in the right-hand panels are for the short segments specified
by the bars at left.

The TRACE mission is described by
	\citet{1999SoPh..187..229H}. 
We use ultraviolet image sequences downloaded from the TRACE
archive\footnote{\url{http://vestige.lmsal.com/TRACE}}.  They were recorded
on June~1, 2003 at the request of M.~Carlsson (Oslo), who suggested strict
cadence regularity and low data compression in order to minimize
high-frequency artifacts, in particular those arising from timing
irregularities as analyzed in Sect.~5 of
	\citealias{2001A&A...379.1052K}. 
TRACE was programmed to obtain such image sequences in its 1600-\AA\ and
1700-\AA\ ultraviolet passbands from 8:14 to 18:34~UT.  We selected
uninterrupted subsequences of 1120~images starting at 11:23:12~UT and
ending at 15:25:56~UT.  They have strictly regular cadence at 13~s sampling
interval in both passbands.  The corresponding Nyquist frequency is
$f_\mathrm{Ny}=38.46~\mathrm{mHz}$; the frequency resolution is $\Delta
f=34.34~\mu\mathrm{Hz}$.  The images sample a quiet area of
$256\arcsec\times192\arcsec$ centered at $X=-2.78\arcsec$, $Y=13.46\arcsec$
near the center of the solar disk, corresponding to a field of view of
$512\times384$ square 0.5\arcsec\ pixels.  The 1600-\AA\ and 1700-\AA\
images were alternately exposed for 1.724 and 4.872~s, respectively.  The
mid-exposure delay between the closest pairs of 1600-\AA\ and 1700-\AA\
images is 5.574~s.

The image sequences were processed with the SolarSoft routine
\texttt{trace\_prep} described in the TRACE Analysis
Guide\footnote{\url{http://moat.nascom.nasa.gov/~bentley/guides/tag}}.  It
corrects missing pixels (of which there were none in these data), replaces
saturated pixels with values above 4095, subtracts the dark field, and
corrects for the flat field.  The dark and flat fields were recorded on
November~8, 2001 and March~11, 2003, respectively.  The image brightness
was normalized by the exposure time.

In the course of this analysis it became clear that precise image
alignment, including corrections for differential solar rotation and for
spacecraft pointing jitter, is crucial to Fourier phase-difference analysis
at high frequencies, and that we should considerably improve on the method
used in
	\citealias{2001A&A...379.1052K}. 
In that paper, co-aligned subfields of the 1700-\AA\ sequences
were cross-aligned to the corresponding subfields in the 1600-\AA\
sequences.  This procedure copies alignment errors from one sequence to the
other and so introduces a high-frequency phase-difference signal at the
retardation set by the timing offset between the exposures at the two
wavelengths.  It emerges in Fig.~18 of
	\citealias{2001A&A...379.1052K} 
as a high-frequency drift of the spatially averaged phase-difference curves
towards the average offset caused by non-simultaneous sampling shown in the
center panel of Fig.~28 of
	\citealias{2001A&A...379.1052K}. 
The slower cadence of the October, 14 1998 data also analyzed in
	\citealias{2001A&A...379.1052K} 
caused correspondingly larger offset (Fig.~26).

In the present analysis such erroneous cross-alignment signals are reduced
by significantly improving the alignment procedure.  In order to minimize
the use of interpolation, we measured pointing displacements per image
through an elaborate procedure detailed below and then used these
displacements to resample the original images directly onto an aligned
grid.

We began by shifting every row of each image in
solar $X$ to correct solar rotation including its differential shear,
using the expression of
	\citet{1990SoPh..130..295H}. 
We then aligned each image of 40-image 1600-\AA\ sub-sequences to
the last one of the previous set, comparably to the procedure in
	\citealias{2001A&A...379.1052K}. 
Each 1700-\AA\ image was subsequently cross-aligned to the corresponding
coarsely aligned 1600-\AA\ image taken 5.574~s before.  We then applied
spatial smoothing through $5\times5$ pixel boxcar averaging to every image,
and, merging the two sequences, applied temporal smoothing per pixel by an
eighteen-image boxcar average.  Alignment of each individual image of the
de-rotated sequences to this smoothed average yields displacement vectors
per image.  SolarSoft routine \texttt{tr\_get\_disp} was employed in all
alignment computations.

\figureone{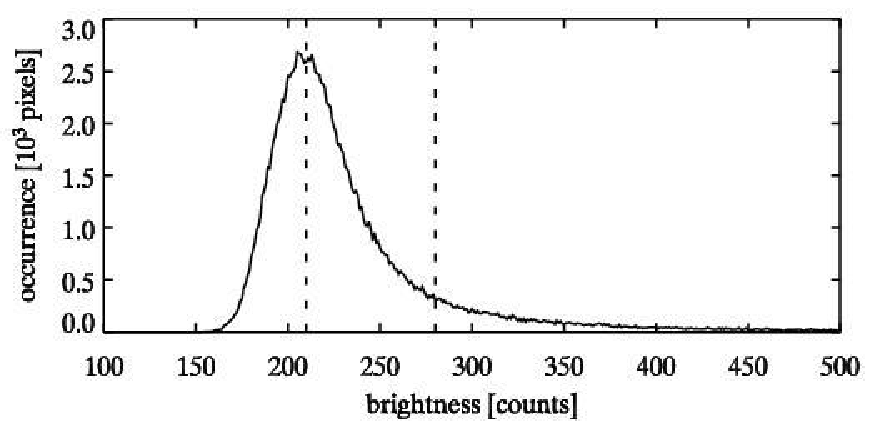}{fig:avghist}{Brightness histogram of one
80-minute average of the 1600-\AA\ sequence.  The dotted lines define the
split between network (right), intermediate (middle), and internetwork
(left).}

\figuretwo{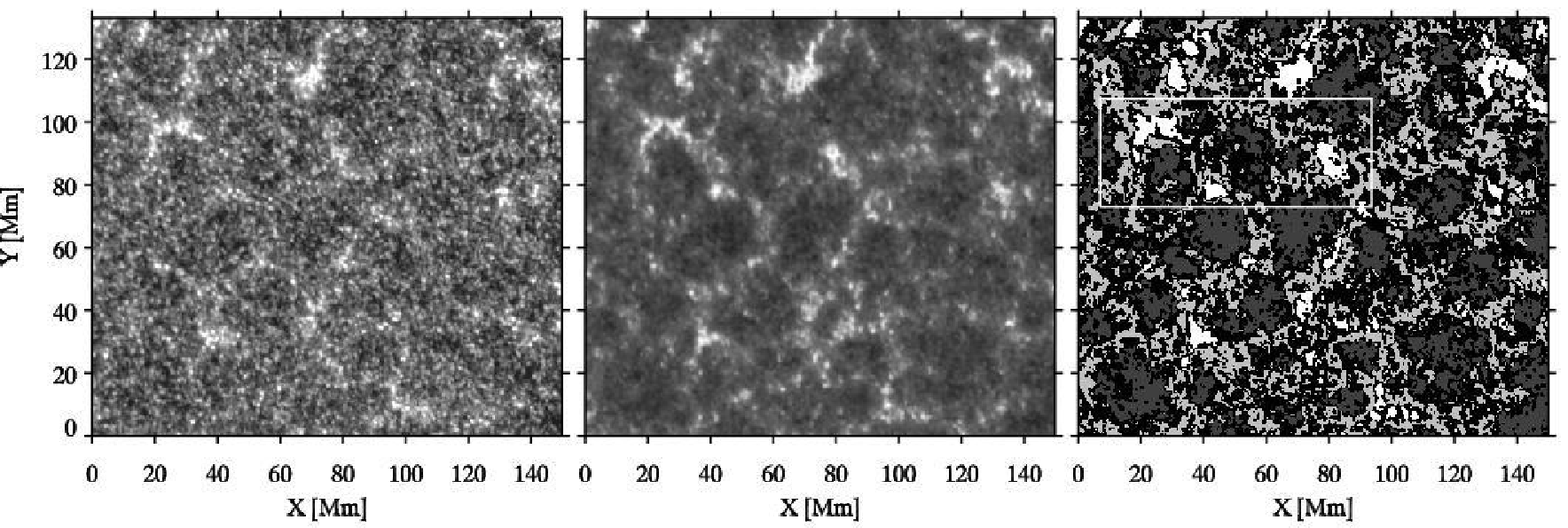}{fig:masks}{\emph{First panel}: sample image from
the 1600-\AA\ sequence taken at 11:32:22~UT.  The intensity was clipped and
scaled logarithmically in order to gain contrast in the internetwork.
\emph{Second panel}: 80-minute 1600-\AA\ average (12:44\,--\,14:05~UT)
using the same gray scale.  \emph{Third panel}: the pixel masks applied to
all images in Sect.~\ref{sec:confusograms}.  Dark gray, light gray, and
white respectively denote internetwork (28413~pixels), intermediate
(36914~pixels), and network (7430~pixels).  Black pixels are discarded.
The box specifies the subfield selected for the power maps in
Fig.~\ref{fig:powermosaic}.}

Figure~\ref{fig:disp1600} shows the resulting displacement corrections for
the 1600-\AA\ sequence.  These are the frame-by-frame residuals after the
initial correction for differential rotation.  They primarily describe
pointing errors.  Both $X$ and $Y$ components show oscillatory behavior
with about 1.5-pixel amplitude and approximately 100-minute periodicity
caused by the spacecraft's orbital motion.  The enlargements at right show
ragged excursions with quarter-pixel amplitudes on short timescales which
reflect pointing jitter.  Solar rotation causes a much larger additional
drift in the horizontal direction.  The de-rotation correction ranges from
$-73.33~\mathrm{pixels}$ at the equator to $-72.83~\mathrm{pixels}$ at the
bottom of the field of view.  Our use of whole-field alignment
automatically corrects for any departures from the initially applied
rotation law except for those in differential shear.  The error estimates
of
	\citet{1990SoPh..130..295H} 
suggest that the remaining shear errors are within $0.004~\mathrm{pixel}$
over our range in solar $Y$ and time.

In the final step of the alignment procedure, the sophisticated algorithm
described by
	\citet{2004SoPh..219....3D} 
is used to re-sample the original images onto a
$432\times384~\mathrm{pixel}$ grid corrected for differential solar
rotation, for spacecraft orbital motion and jitter, and for the re-mapping
from planar to spherical coordinates.  The area of incomplete sampling due
to solar rotation is discarded, as is a vertical strip at the left-hand
edge of the field of view which erroneously appears bright in one 1700-\AA\
image.  The resulting images consist of $432\times384$
$0.348~\mathrm{Mm}$-square pixels.

For part of our analysis, i.e., the spatially averaged Fourier spectra
presented in Fig.~\ref{fig:triconfuse} in Sect.~\ref{sec:confusograms}, we
followed the procedure of
	\citealias{2001A&A...379.1052K} 
to divide the field of view into internetwork, intermediate, and network
areas through classification of the time-averaged 1600-\AA\ brightness per
pixel.  Temporal averaging increases the contrast between the rapidly
changing internetwork brightness and the more stable network emission.  The
1600-\AA\ sequence was split into three parts of approximately 80~minutes
duration.  Figure~\ref{fig:avghist} displays the brightness distribution
after averaging over one 80-minute part.  It has a Gaussian peak and an
extended high-brightness tail.  A pixel is classified as internetwork if in
all three 80-minute averages its brightness remains below the left-hand
dotted line, which is chosen near the three peaks.  Pixels with brightness
above the right-hand dotted cutoff in all three averages are classified as
network.  Pixels that fall between the two lines in all three averages are
classified as intermediate.  Pixels that change category between averages
are discarded.  This category amounts to 56\% of all pixels due to the long
sequence duration.  It is large to avoid any mixing of internetwork,
intermediate, and network behaviour.

Particle hits were not corrected by interpolation but were removed on the
basis of their high-frequency signature for the analysis in
Sect.~\ref{sec:confusograms}.  Their single-image appearance produces
anomalously strong high-frequency power.  We applied a spatial mask to
remove all pixels as well as their immediate neighbors that show Fourier
power in excess of three times the average in the highest 50 frequency bins
(36.7\,--\,38.5~mHz).

Figure~\ref{fig:masks} presents a sample 1600-\AA\ image, one of the three
80-minute averages, and the three masks.

\section{Analysis and results}\label{sec:analysis}

Fourier power, coherence and cross-power spectra were computed per pixel in
the 1600-\AA\ and 1700-\AA\ image sequences over their full 243-minute
duration following the recipes in Sect.~3 of
	\citealias{2001A&A...379.1052K}. 
We follow
	\citealias{2001A&A...379.1052K} 
also in the presentation of the resulting power maps, spatially-averaged
temporal Fourier spectra, and two-dimensional \kf\ diagrams.  Here, the
emphasis is on high-frequency behavior and its significance.

\subsection{Spatially resolved Fourier power maps} \label{sec:powermaps}
As in 
	\citealias{2001A&A...379.1052K} 
we distinguish three different normalization choices in displaying Fourier
power per pixel as spatially resolved maps, namely plotting the
non-normalized oscillatory energy itself (``power''),
\begin{equation}
	P_E(x,y,f)=|\mathcal{I}(x,y,f)|^2\,,
\end{equation}
the fractional modulation signal obtained by dividing the oscillatory
energy by the zero-frequency power (``modulation''),
\begin{equation}
	P_f(x,y,f)=\frac{|\mathcal{I}(x,y,f)|^2}{|\mathcal{I}(x,y,0)|^2}\,,
\end{equation}
and ``Leahy'' normalization obtained by dividing the energy by the
zero-frequency amplitude,
\begin{equation}
	P_\mathrm{L}(x,y,f)=\frac{|\mathcal{I}(x,y,f)|^2}{|\mathcal{I}(x,y,0)|}\,,
\end{equation}
where $x$ and $y$ are spatial coordinates, $f$ is the temporal frequency,
and $\mathcal{I}(x,y,f)$ denotes the Fourier transform of the intensity
measured by TRACE at location $(x,y)$ at frequency $f$.  Leahy
normalization is used in the literature to estimate power-peak significance
(e.g.,
   \citealp{1983ApJ...266..160L}, 
   \cite{1999A&A...347..335D}) 
but was not used in
	\citealias{2001A&A...379.1052K}. 

We here add 95\% significance estimation following
	\citet{1975ApJS...29..285G} 
and first compare this to Fisher's method of randomization described by,
e.g.,
	\citet{Bradley1968} 
and 
	\citet{1985AJ.....90.2317L} 
and applied to solar data by, e.g., 
	\citet{2001A&A...368.1095O} 
and
	\citet{2003A&A...401..685M}. 
Its assumption is that there is no signal at any frequency, so that the
temporal order in which the data were taken becomes irrelevant.  Comparison
of the actual power spectrum to the spectra of temporal permutations of the
data sequence then yields a significance estimate.  The test is used
iteratively, progressively deleting significant peaks until no new peaks
are found.  It puts no constraint on the noise power distribution at any
given frequency, but the assumption that all samples are temporally
uncorrelated implies frequency-independent white noise.  For large data
sets it becomes impractical to repeatedly compute all possible
permutations.  Actual tests are therefore usually limited to a few hundred
permutations, but even then remain computationally expensive.

The much simpler significance estimation of
	\citet{1975ApJS...29..285G} 
assumes that at any frequency the real and imaginary parts of the Fourier
power have independent normal distributions.  It requires explicit
specification of the noise power as a function of frequency, i.e., the
noise is not assumed to be white.

\figureone{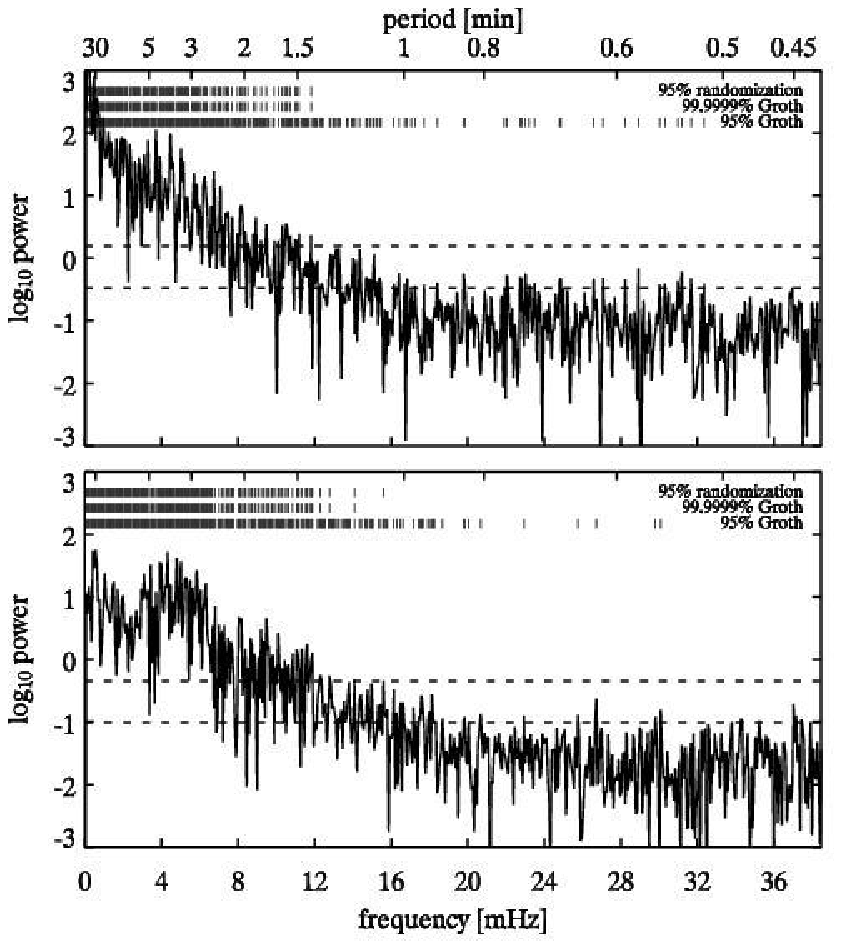}{fig:randplot}{Comparison of the randomization
test to significance estimation following
	\citet{1975ApJS...29..285G}. 
\emph{Upper panel}: network pixel.  \emph{Lower panel}: internetwork pixel.
The ragged curves are the temporal Fourier power at 1600~\AA\ using TRACE
data units divided by the exposure time, on logarithmic scales.  In each
panel, the top row of tick marks identifies all significant peaks according
to the randomization test.  The second and third rows identify significant
peaks at the 99.9999\% and 95\% significance levels using Groth's test
assuming white noise and absence of signal above $f=24~\mathrm{mHz}$.  The
dotted lines show the corresponding cutoff levels.}

We compare the randomization test with Groth's test in
Fig.~\ref{fig:randplot} for an internetwork and a network pixel, adopting
95\% confidence levels in both tests.  In the randomization test, a power
peak that is above the maximum power of the randomized data in more than
95\% of 500 permutations is considered statistically significant.  Such
peaks are subsequently removed in the iterative re-application of this
procedure until no more significant peaks are found.  For Groth's test we
decided from visual inspection to assume that the power spectra display
white noise above $f=24~\mathrm{mHz}$.  The corresponding 95\% significance
cutoff is 2.996~times higher than this noise level.

The top and bottom rows of tick marks in Fig.~\ref{fig:randplot} specify
the positions of the peaks that are estimated to be significant by the two
methods.  It is obvious that the randomization test is far more rigorous
than Groth's test, as pointed out by 
	\citet{2003A&A...401..685M}. 
The middle rows of ticks result when the upper dotted line is used as
Groth-test cutoff level, at fourteen times the noise level corresponding to
99.9999\% confidence.  It closely matches the peak-finding by the
randomization test.  Thus, for these data a stringent Groth test may
replace the randomization test at much smaller computational cost.  This is
likely to hold for other data with white noise.

The network pixel in the upper panel of Fig.~\ref{fig:randplot} has larger
low-frequency signal than the internetwork pixel in the lower panel, a
different power hump around 5-min periodicity, and higher high-frequency
noise but rather similar peak survival above the lenient 95\% Groth cutoff
estimate.

\figuretwo{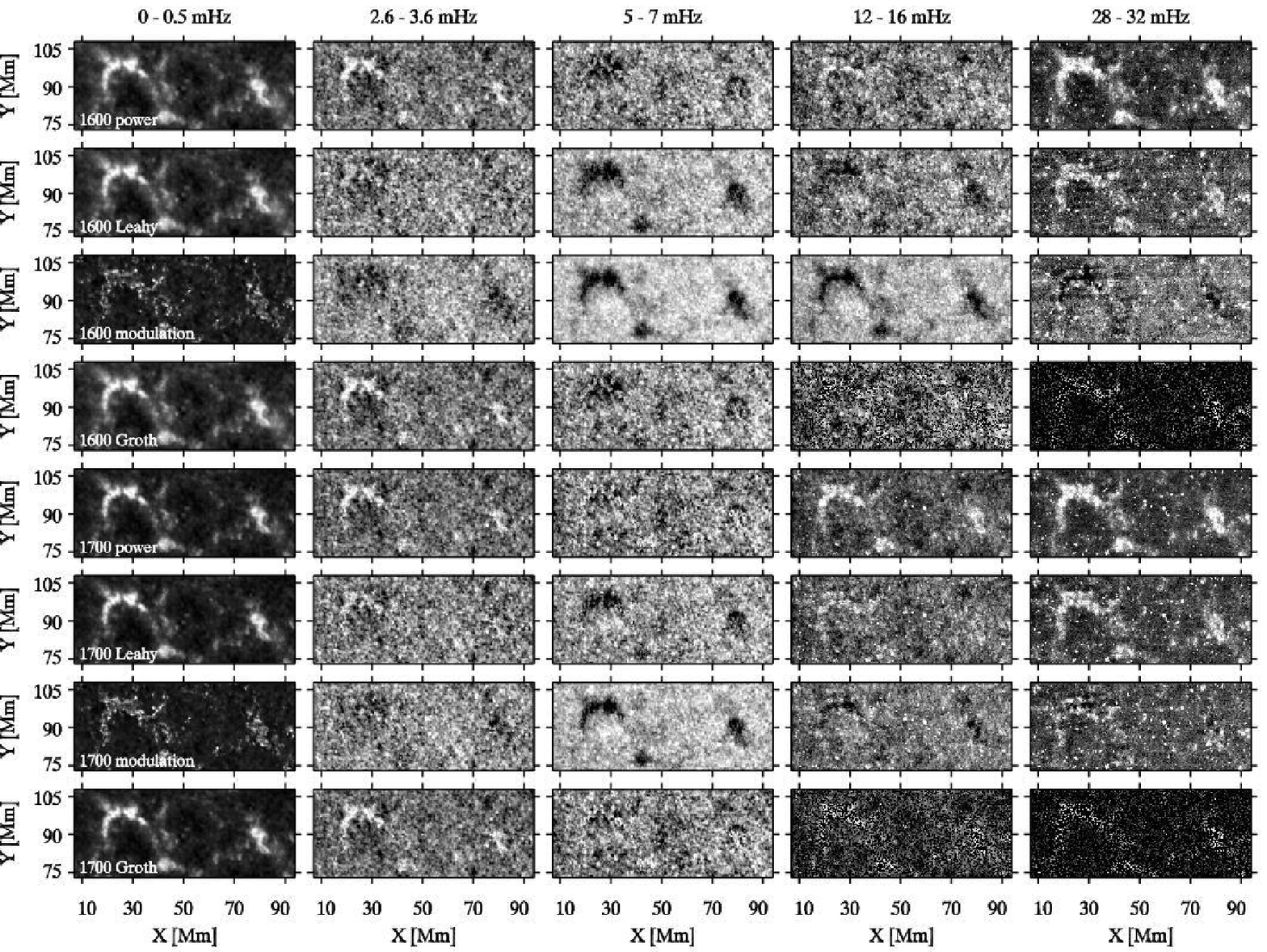}{fig:powermosaic}{Spatially resolved power
maps using different methods of normalization for the subfield shown in
Fig.~\ref{fig:masks}.  The grayscale displays the logarithm of the temporal
Fourier power, clipped to improve contrast.  \emph{Columns}: different
frequency bands as specified at the top.  \emph{Rows}: 1600-\AA\ and
1700-\AA\ passbands, with different power normalization as specified in the
first column.}

Figure~\ref{fig:powermosaic} expands such comparison of internetwork versus
network by displaying spatial power maps for the small but illustrative
subfield specified by the rectangle in Fig.~\ref{fig:masks} for both the
1600-\AA\ and 1700-\AA\ passbands.  The subfield is shown in different
temporal frequency bands, with the three different normalizations
(power, Leahy, and modulation, respectively), and finally also
without normalization but with all pixels having power below the 95\% Groth
cutoff made black.

The first column shows low-frequency power.  It shows the stable nature of
bright network.  The power-normalized modulation maps show noisy behavior
from the division because the frequency band is close to zero Hz.

The second-column frequency range of 2.6\,--\,3.6~mHz corresponds to
periodicities around 5~minutes.  The network appears power-bright in the
unnormalized maps, about equal to the internetwork in the
amplitude-normalized Leahy maps, and power-dark in the modulation maps.
Thus, the choice of normalization affects the apparent relative dominance
of network and internetwork oscillations, as discussed extensively in
	\citealias{2001A&A...379.1052K}. 
Note that in all representations a power-dark moat appears around the
network.

The 5\,--\,7~mHz maps describe the chromospheric
three-minute oscillation which pervades internetwork areas
  (e.g., \cite{Rutten1995b}). 
They indeed show the network power-dark in all representations.  There are
irregular power-bright ``aureole'' patches near network (cf.
	\citealias{2001A&A...379.1052K}). 

The two high-frequency columns on which we concentrate here illustrate the
care that must be taken in interpreting such power maps.  In the rightmost
column (28\,--\,32~mHz) the network stands out very brightly in the
unnormalized power maps, inviting a claim that the magnetic elements making
up the network display high-frequency wave heating.  On the other hand, the
network appears power-dark in the modulation maps, inviting a claim that
high-frequency oscillations are suppressed in magnetic elements.  However,
the close spatial correspondence of both these bright and dark features
with the bright network in the unnormalized maps in the first column
suggests strongly that they are simply due to the larger overall network
brightness.  A similar power-contrast flip is seen in the 12\,--\,16~mHz
maps for the 1700-\AA\ passband, but the unnormalized 1600-\AA\ map for
these frequencies appears rather featureless.  The latter copies directly
into the Groth map, but with considerable pixel deletion wherever the power
averaged over the frequency range drops below the cutoff.  In the rightmost
column the Groth maps accept only a minor fraction of the pixels as
significant, both for the network and the internetwork.  A more strict
criterion, and certainly the 95\% iterative randomization test, would
reject all.  The patterns seen in the other 28\,--\,32~mHz maps thus are
most likely artifacts caused by sources of high-frequency errors with some
sensitivity to the low-frequency power.  Note that Leahy normalization
diminishes the apparent structure for 1700~\AA\ but turns it power-dark at
1600~\AA.

The bright specks in the 28\,--\,32~mHz maps are due to particle hits.
They produce high-frequency signal through their instantaneous appearance.
The pattern of horizontal stripes results from TRACE's JPEG data
compression (K.~Muglach, private communication).  It appears as a grid
pattern with 8-pixel mesh size in comparable power maps computed from the
original non-aligned image sequences.  The compensation for solar rotation
smears out the vertical grid components, leaving only the horizontal ones.

\subsection{Spatially averaged Fourier power, phase difference and
coherence}\label{sec:confusograms}
We now turn to temporal Fourier analysis with spatial averaging over
the different pixel categories defined by the third panel of
Fig.~\ref{fig:masks}.  The averaging is performed on the
Fourier measurements per pixel.  It reduces the noise in these
measurements and therefore improves the detection of relatively small
modulation signals.  Following
	\citealias{2001A&A...379.1052K}, 
the power and coherence are averaged directly over all relevant pixels.
In the present analysis we compute coherence per pixel using frequency
smoothing over 9 bins rather than 5.

The phase differences are again averaged with cross-power weighting as
introduced by
	\citet{1979ApJ...231..570L}, 
i.e., the spatial average over the phase differences of all pixels
transmitted by the mask per frequency bin is defined as the angle of the
vector sum of the cross-powers of all contributing pixels with reference to
the real axis.  The advantage of such vector averaging is that it makes
signals stand out even in the presence of much larger noise.  For pure
noise the vector mean does not go to zero or some other definite value but
fluctuates randomly over the full $-180$ to $+180$~degree range between
adjacent frequency bins.  A small signal, much smaller than the noise, may
therefore emerge as a systematic pattern across multiple bins.

\figureone{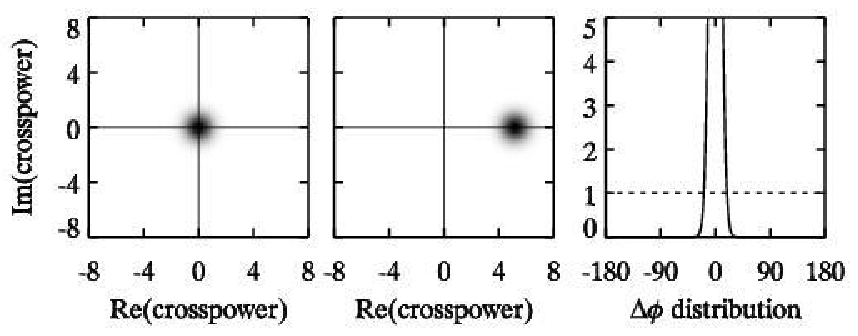}{fig:cpnoise}{Illustration of vector-averaging
phase differences for 30\,000~pixels. \emph{First panel}: distribution of
the cross-power vector sum for pure Gaussian noise in the complex plane.
\emph{Second panel}: same as the first panel, but with a signal with
amplitude of only 3\% of the rms noise with 0~degree phase difference
added.  The vector summation of the 30\,000 samples shifts the scattercloud
significantly to the right.  \emph{Third panel}: corresponding
phase-difference distributions for pure noise (dashed, flat) and with the
signal added (solid, peaked).  The latter reaches 18.4 at phase difference
0 with 20-degree half-width.}

This is illustrated by Fig.~\ref{fig:cpnoise} which displays
simulation results for vector-averaged cross-power
distributions of pure noise (first panel) and of pure noise with a
much smaller superimposed signal with fixed phase difference (second
panel).  In the latter case, the vector addition over 30\,000 pixels
with small but systematic signal shifts the much wider scattercloud to
a location well separated from the origin, making the phase-difference
distribution in the third panel strongly peaked.

\figuretwo{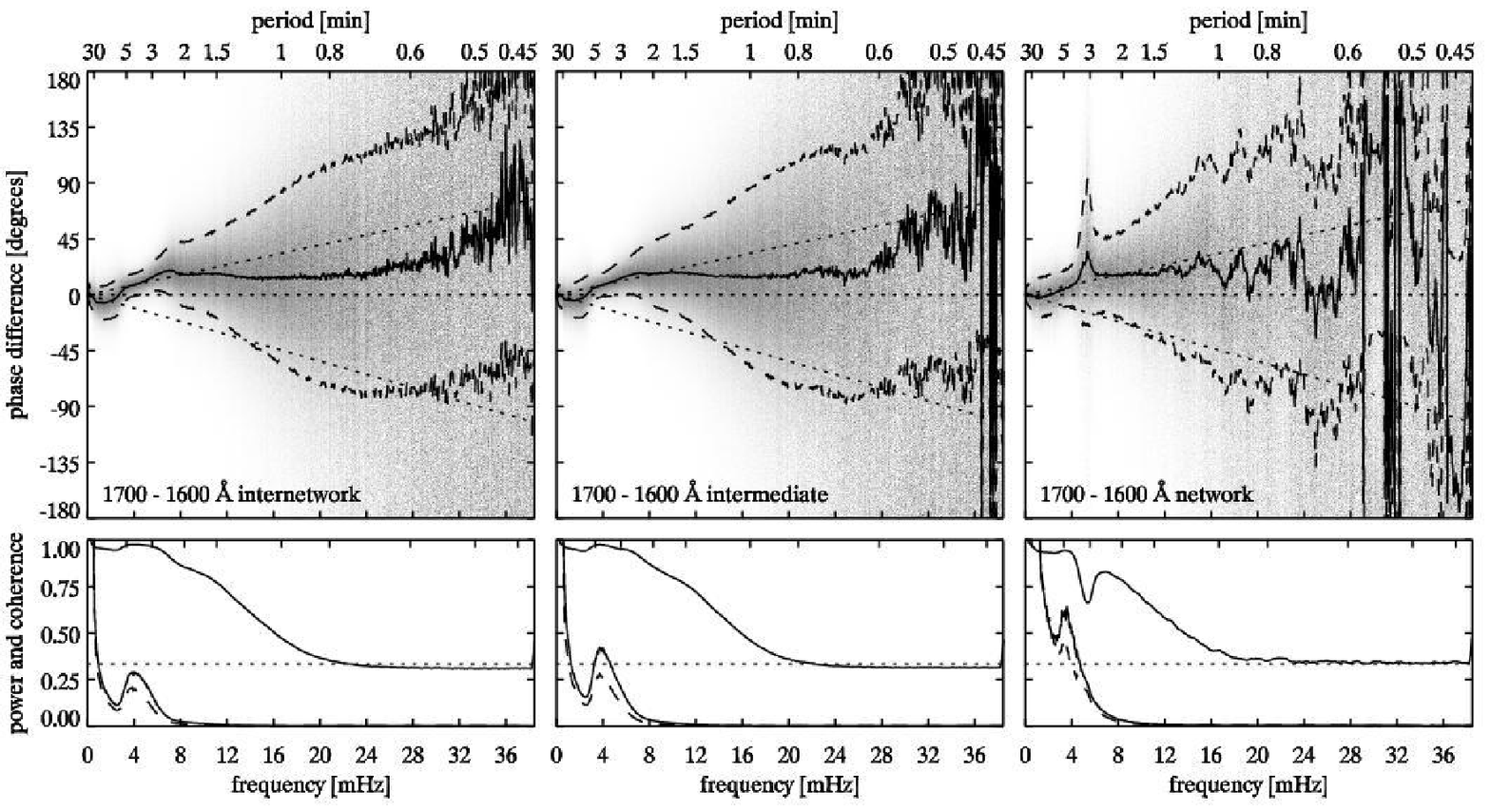}{fig:triconfuse}{Temporal Fourier spectra,
spatially averaged over internetwork (left), intermediate
(middle), and network (right), plotted against
frequency up to the Nyquist limit with the corresponding periodicities
shown along the top.  The format corresponds to Figs.~18\,--\,19 of
	\citealias{2001A&A...379.1052K}, 
adding dotted lines indicating the sampling time delays and omitting
1$\sigma$ rms estimates for power and coherency to avoid clutter.
\emph{Upper panels}: phase-difference spectra.  \emph{Lower panels}:
coherence (upper curves) and power spectra (solid for 1600~\AA, dashed for
1700~\AA).  The random-noise estimate for the coherence is $C=0.33$.  The
power spectra are on linear scales and are all scaled by the same factor.}

Figure~\ref{fig:triconfuse} presents results from the new TRACE sequences
in the form of power, coherence, and phase-difference spectra with spatial
averaging separated between the internetwork, intermediate, and network
areas.  The format is similar to Figs.~18\,--\,22 of
	\citealias{2001A&A...379.1052K} 
but adds two dotted lines in the phase-difference panels.  These are the
phase shifts associated with the temporal sampling delay due to the
non-simultaneous exposures in the two passbands.  We have aligned the two
sequences to match the closest pair combinations, corresponding to the
upper dotted lines.  All of our phase-difference evaluations are corrected
for this sampling offset, which means that signals that are intrinsically
in phase in the two passbands should indeed end up along the horizontal
$\Delta\phi=0$ axis.  Measurements that end up on a dotted line imply
modulation with phase delay exactly matching the corresponding sampling
delay.  The grayscaled scattercloud represents individual pixels.  In the
case of pure noise the $1\sigma$ rms estimates cover 68\% of the full
figure height around a randomly fluctuating mean.

The results in Fig.~\ref{fig:triconfuse} are similar to those in Fig.~18 of
	\citealias{2001A&A...379.1052K} 
except for the high-frequency phase differences of interest here.  The
present results are more reliable thanks to the regular sampling cadences,
lower data compression, and better image alignment.

The internetwork phase differences reach a wide maximum at
$f\approx7~\mathrm{mHz}$ and then remain well defined at positive values up
to the Nyquist frequency, but with increasing noise above 20~mHz.  There is
no drop to negative values as in
	\citealias{2001A&A...379.1052K}, 
which we now attribute to the cross-alignment used there as discussed in
Sect.~\ref{sec:observations} above.  However, the present results also show
a drift to the phase difference associated with the timing delay at high
frequencies.  The internetwork power spectra show acoustic humps around
4~minutes and become negligible above 12~mHz.  The coherence also peaks
around 4~mHz and drops to the 9-bin noise level near 20~mHz.

The network phase differences in the third column of
Fig.~\ref{fig:triconfuse} are much noisier due to the far smaller number of
pixels.  Nevertheless, they show a narrow peak of increased phase
difference and reduced coherence around three-minute periodicity which is
not present in Fig.~18 of
	\citealias{2001A&A...379.1052K} 
and which we deem significant.  At high frequencies they become more
erratic and shift to the timing correction line, which points to a
systematic error.

The intermediate-class pixels in the center column of
Fig.~\ref{fig:triconfuse} produce primarily internetwork-like behavior.

\subsection{Two-dimensional Fourier power and phase difference}
\label{sec:komega}

Figure~\ref{fig:komerged} presents two-dimensional Fourier power and phase
difference spectra in the form of \kf\ diagrams.  They mix the network,
internetwork and intermediate areas.  Particle hits were not removed
because their contribution to the noise is small except at high spatial and
temporal frequencies where the diagrams are noisy anyhow.

The power and phase differences are averaged over rings of constant
$k_\mathrm{h}$, with $k_\mathrm{h}^2=k_\mathrm{x}^2+k_\mathrm{y}^2$,
assuming absence of preferred horizontal propagation directions.  The
number of samples per ring increases with $k_\mathrm{h}$ up to the Nyquist
frequency per axis $k_\mathrm{x,Ny} =
k_\mathrm{y,Ny}=9.0~\mathrm{Mm}^{-1}$.  Beyond this value, $k_\mathrm{h}$
can still be computed but with fewer samples and increasing loss of
isotropy in each successive bin, up to
$k_\mathrm{h}=\sqrt{2}\,k_\mathrm{x,Ny}=12.8~\mathrm{Mm}^{-1}$ which
samples only oblique propagation.

The left-hand panel of Fig.~\ref{fig:komerged} shows the \kf\ diagram for
1600-\AA\ power.  The acoustic $p$-mode ridges and pseudo-ridges above the
Lamb line were extensively discussed in
	\citealias{2001A&A...379.1052K}. 
There is no particular structure evident in the high-frequency regime of
interest here.  At low frequencies there is a ridge of enhanced power at
high spatial wavenumbers, approximately corresponding to
$f=(1/2\pi)\,2~\mathrm{km\,s}^{-1}$ which is caused by the compensation for
solar rotation.  Features that are fixed to the CCD camera, such as the
results of an imperfect flat field, or ``hot'' pixels, move apparently with
this speed against the direction of solar rotation and produce power.

\figurethree{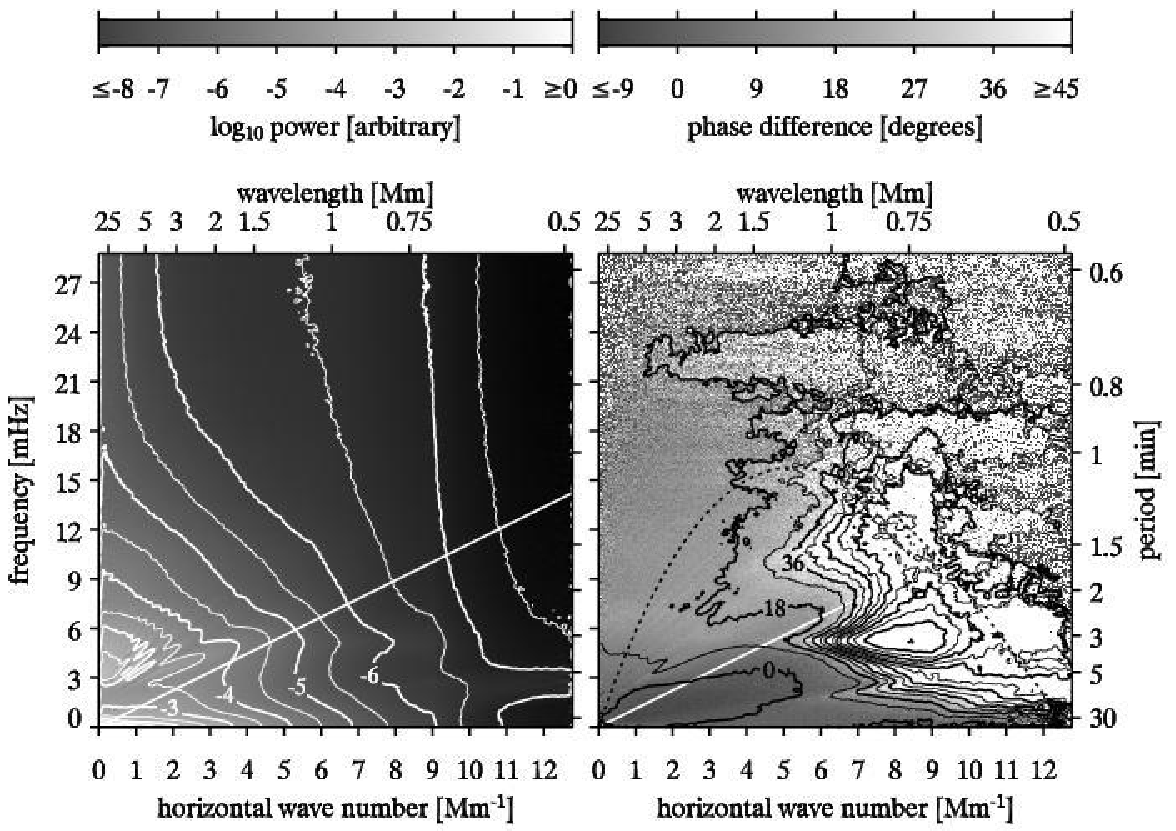}{fig:komerged}{\emph{Left}: power in the
1600-\AA\ image sequence plotted as function of horizontal wave number
$k_\mathrm{h}$ and temporal frequency $f$.  The logarithmic grayscale is
clipped to show the ridges around $1~\mathrm{Mm}^{-1}$ and
$5~\mathrm{mHz}$.  The slanted line is the Lamb line
$f=(1/2\pi)\,c_\mathrm{s}k_\mathrm{h}$ with
$c_\mathrm{s}=7~\mathrm{km\,s}^{-1}$.  \emph{Right}: corresponding phase
differences between the 1700-\AA\ and 1600-\AA\ image sequences.  To avoid
figure clutter, contours are only shown if they lie below the
dashed curve or enclose a large area.  The grayscale is clipped at $-9$
and $45~\mathrm{degrees}$ to increase contrast.  The white blob peaking at
$k_\mathrm{h}=9~\mathrm{Mm}^{-1}$ and $f=5~\mathrm{mHz}$ reaches 110-degree
difference.  The pepper-and-salt regions reflect noise.}

The right-hand panel of Fig.~\ref{fig:komerged} displays the corresponding
\kf\ diagram for phase difference between the 1700-\AA\ and 1600-\AA\
sequences.  The acoustic ridges stand out through larger phase difference,
as discussed in
	\citealias{2001A&A...379.1052K}. 
The wedge of negative phase difference at low frequencies and wave numbers
was attributed to atmospheric gravity waves by
	\citet{2003A&A...407..735R}. 
The effects of solar rotation are again visible as a ridge of slightly
increased phase difference.  We attribute this increase to the systematic
$(-0.285, 0.273)$-pixel image offset between the two passbands before
alignment, which causes apparently traveling features fixed to the CCD to
appear at a given solar location with some time delay in the two passbands.

At the highest temporal frequencies of interest here, noise is easily
identified as pepper-and-salt patterning where the phase differences jump
widely from one bin to the next.  A large patch of smooth variation extends
up to about 20~mHz in frequency and $4~\mathrm{Mm}^{-1}$ in wavenumber.
This patch contributes most to the definite phase behavior in
Fig.~\ref{fig:triconfuse}, which is therefore set by these spatial scales.

The conspicuous white blob of large positive phase difference located at
$0.75~\mathrm{Mm}$ wavelength and 3-minute periodicity, with an extended
tail upward, is enigmatic.  The left-hand diagram suggests enhanced power at
this location.  It seems likely that the peak in the network panel of
Fig.~\ref{fig:triconfuse} corresponds to this blob, and that therefore the
source should be sought in the network.  It is very tempting to attribute
it to solar three-minute waves with small horizontal extent in magnetic
elements.  Its exceedingly large value (up to 110~degree difference at its
center), the clear reduction in coherence, and the lack of such a blob in
comparable phase-difference diagrams sampling the \CaIIH\ line core and
inner wing from observations made with the Dutch Open Telescope
(unpublished analysis analogous to Fig.~7 of
  \cite{2004A&A...416..333R}), 
would then suggest phenomena in the transition region that
leave a signature in these data through the \CIV\ doublet at $\lambda
= 1548~\mbox{\AA}$ and $1550~\mbox{\AA}$
   (cf.\ \cite{1998SoPh..183...29H}). 
However, the blob lies at the spatial Nyquist frequency per
horizontal axis, and its shape varies with changes in the image alignment
procedure.  We reluctantly conclude that the blob is likely a TRACE
artifact, presumably of instrumental origin, introduced by the data
processing, or a combination of these.

\section{Discussion}\label{sec:discussion}

We find intriguing high-frequency behavior in all our Fourier displays, but
at the same time also find reasons to disbelieve these signatures above
20~mHz or even lower frequencies.  The pixel-by-pixel power maps in
Fig.~\ref{fig:powermosaic} have disconcerting high-frequency contrast
sensitivity to the type of normalization above 10~mHz.  The spatial
averaging in Figs.~\ref{fig:triconfuse} and~\ref{fig:komerged},
respectively over mask-selected pixel types and annuli, improves the
significance in phase-difference measurement, but the high-frequency
behavior in Fig.~\ref{fig:triconfuse} is puzzling in the trends towards
the timing delay lines and the absence of purely random behavior even at
the highest frequencies.  The prominent white blob in the phase-difference
panel of Fig.~\ref{fig:komerged} is presumably an artifact.

The phase-difference averaging with cross-power weighting over the
different pixel categories employed in Sect.~\ref{sec:confusograms} is the
most sensitive method to identify weak oscillation signatures in the
presence of noise.  Each phase-difference diagram in
Fig.~\ref{fig:triconfuse} indicates systematic non-random behavior out to
frequencies far beyond the extent of measurable power or even of measurable
coherence.  This was already the case in Figs.~18\,--\,19 of
	\citealias{2001A&A...379.1052K} 
and also in the similar diagrams from groundbased \CaIIH\ spectrometry in
Figs.~20\,--\,22 of
	\citealias{2001A&A...379.1052K}. 
It is attractive to believe that the cross-power weighting indeed enhances
the sensitivity of the phase-difference measurement to very small signals
otherwise drowned in noise out to well above the coherence limit of at most
20~mHz, but it is alarming that even at the highest frequencies the phase
differences do not show the randomness expected, and that in all three
panels in Fig.~\ref{fig:triconfuse} it seems to favor the instrumental
timing correction.  It is likely that residual image alignment errors are
the cause of this anomalous behavior.

It is well-known that the increasing lack of response due to wide
contribution functions
 (e.g.,
  \cite{1975SoPh...43..289B}, 
  \cite{1976A&A....51..189D}, 
  \cite{1980A&A....84...99S}, 
  \cite{1980A&A....91..251D}) 
hampers the detection of high-frequency signals.  It was recently
elaborated in TRACE context by
  \citet{Fossum2004}. 
In addition, we have learned from M.~Carlsson (private communication) that
simulations of acoustic waves propagating upward in the solar atmosphere as
in the well-known \CaII\ \HtwoV\ grain simulation of
  \citet{1997ApJ...481..500C} 
subjected to computational 1700\,--\,1600-\AA\ phase-difference analysis
which emulates the observational analysis presented here meets unexpected
computational problems at low signal to noise and high frequencies.

On the other hand, we have reproduced our phase-difference results 
in tests using double precision computation.  Very similar non-random
positive phase-difference behavior also appears up to 20~mHz in Fig.~20 of
	\citealias{2001A&A...379.1052K}, 
based on the \CaIIH\ slit spectrometry of
	\citet{1993ApJ...414..345L} 
and measured from \CaIIH\ wing intensities and \FeI\ blend Doppler shifts
formed at lower and similar heights as the ultraviolet continua sampled by
TRACE.  
The comparable signature in \IminI\ and \VminV\ diagnostics with
negative \VminI\ lag shown there is in agreement with acoustic waves.  
The steep \VminV\ signature of upward propagation in
internetwork areas present in the lower-left panel of Fig.~21 of
	\citealias{2001A&A...379.1052K} 
seems significant also.

In summary, the coherence spectra in Fig.~\ref{fig:triconfuse}, the close
agreement of the phase differences in that figure with those from \CaIIH\
spectrometry in
	\citealias{2001A&A...379.1052K}, 
and the smoothness of the corresponding gray area in
Fig.~\ref{fig:komerged}, all taken together lead us to believe that the
phase-difference signals derived from these new TRACE sequences have a
solar origin up to 20~mHz at least in the internetwork, and are to be
attributed to acoustic waves.

This conclusion supports the detection of high-frequency acoustic waves by
  \citet{2002A&A...395L..51W} 
as significant Doppler-shift power in the 10\,--\,20~mHz frequency band from
differential \FeI\ measurements addressing similar atmospheric heights.
The drop of power with frequency in our Figs.~\ref{fig:randplot},
\ref{fig:triconfuse} and \ref{fig:komerged} suggests that their detection
is dominated by the lower frequencies in this band.  Our results indicate
wave presence also at the higher frequencies.

The ultraviolet continua used here suffer from strong scattering while the
TRACE filter bandwidths are wide and overlap considerably.  Numerical
simulations as those presently underway at Oslo may explain how and why the
phase differences in Figs.~\ref{fig:triconfuse} and \ref{fig:komerged}
level out at positive values.  In our opinion, comparison with detailed
numerical simulations is also required to substantiate any claim that
acoustic waves in the 10\,--\,20~mHz regime compensate fully for the
radiative losses of the chromosphere.

\section{Conclusion}\label{sec:conclusion}

New ultraviolet image sequences from TRACE give evidence of brightness
modulation up to 20~mHz in quiet-sun internetwork.  We interpret this
signal as a signature of acoustic waves.  It is similarly present in
\CaIIH\ and \FeI\ \IminI\ and \VminV\ phase-difference spectra in Fig.~20
of
	\citealias{2001A&A...379.1052K} 
and it supports the detection of acoustic wave power in the
10\,--\,20~mHz frequency band from \FeI\ Doppler-shift measurements by
  \citet{2002A&A...395L..51W}. 
The evidence for modulation at higher frequencies remains
inconclusive.

TRACE-like ultraviolet imaging will be achieved with the Atmospheric
Imaging Assembly on NASA's Solar Dynamics Observatory, but it is not yet
clear whether its hardware and operation will permit better high-frequency
modulation measurement than with TRACE.  New ground-based telescope
technology, in particular large aperture combined with adaptive optics,
will provide accurate Doppler shifts from integral field spectroscopy at
high cadence and low noise of the same layers using appropriate spectral
lines in the optical.  Numerical simulations may contribute quantification
of the corresponding energy budgets.

\acknowledgements We thank M.~Carlsson for suggesting these TRACE observations
to the third author and for sharing simulation insights in the intricacies of
phase-difference determination with the second author.  We also thank
C.E.~DeForest, J.~Leenaarts, C.C.~Kankelborg, J.M.~Krijger, B.W.~Lites,
K.~Muglach and R.A.~Shine for advice and discussions, and the referee
for suggesting many clarifications.  A.G. de Wijn and R.J. Rutten acknowledge
travel support from NASA (contract NAS5-38099) and the Leids Kerkhoven-Bosscha
Fonds, and are indebted to the Lockheed Martin Solar and Astrophysics Lab.\ at
Palo Alto, the solar physics group of Montana State University at Bozeman, and
the High Altitude Observatory at Boulder for hospitality.



\begin{thebibliography}{27}
\expandafter\ifx\csname natexlab\endcsname\relax\def\natexlab#1{#1}\fi

\bibitem[{{Beckers} \& {Milkey}(1975)}]{1975SoPh...43..289B}
{Beckers}, J.~M. \& {Milkey}, R.~W. 1975, \solphys, 43, 289

\bibitem[{{Biermann}(1948)}]{1948ZA.....25..161B}
{Biermann}, L. 1948, 25, 161

\bibitem[{Bradley(1968)}]{Bradley1968}
Bradley, J.~V. 1968, Distribution-free statistical tests (Prentice-Hall)

\bibitem[{{Carlsson} \& {Stein}(1997)}]{1997ApJ...481..500C}
{Carlsson}, M. \& {Stein}, R.~F. 1997, \apj, 481, 500

\bibitem[{{DeForest}(2004)}]{2004SoPh..219....3D}
{DeForest}, C.~E. 2004, \solphys, 219, 3

\bibitem[{{Deubner}(1976)}]{1976A&A....51..189D}
{Deubner}, F.-L. 1976, \aap, 51, 189

\bibitem[{{Doyle} {et~al.}(1999){Doyle}, {van den Oord}, {O'Shea}, \&
  {Banerjee}}]{1999A&A...347..335D}
{Doyle}, J.~G., {van den Oord}, G.~H.~J., {O'Shea}, E., \& {Banerjee}, D. 1999,
  \aap, 347, 335

\bibitem[{{Durrant}(1980)}]{1980A&A....91..251D}
{Durrant}, C.~J. 1980, \aap, 91, 251

\bibitem[{{Fontenla} {et~al.}(1993){Fontenla}, {Avrett}, \&
  {Loeser}}]{1993ApJ...406..319F}
{Fontenla}, J.~M., {Avrett}, E.~H., \& {Loeser}, R. 1993, \apj, 406, 319

\bibitem[{Fossum \& Carlsson(2004)}]{Fossum2004}
Fossum, A. \& Carlsson, M. 2004, in Waves, Oscillations and Small-Scale
  Transients Events in the Solar Atmosphere: Joint View from SOHO and TRACE,
  Proc.\ of SOHO 13 (ESA Publ.\ Div., ESTEC, Noordwijk: ESA SP--547), 125--129

\bibitem[{{Groth}(1975)}]{1975ApJS...29..285G}
{Groth}, E.~J. 1975, \apjs, 29, 285

\bibitem[{{Handy} {et~al.}(1999){Handy}, {Acton}, {Kankelborg}, {Wolfson},
  {Akin}, {Bruner}, {Caravalho}, {Catura}, {Chevalier}, {Duncan}, {Edwards},
  {Feinstein}, {Freeland}, {Friedlaender}, {Hoffmann}, {Hurlburt}, {Jurcevich},
  {Katz}, {Kelly}, {Lemen}, {Levay}, {Lindgren}, {Mathur}, {Meyer}, {Morrison},
  {Morrison}, {Nightingale}, {Pope}, {Rehse}, {Schrijver}, {Shine}, {Shing},
  {Strong}, {Tarbell}, {Title}, {Torgerson}, {Golub}, {Bookbinder}, {Caldwell},
  {Cheimets}, {Davis}, {Deluca}, {McMullen}, {Warren}, {Amato}, {Fisher},
  {Maldonado}, \& {Parkinson}}]{1999SoPh..187..229H}
{Handy}, B.~N., {Acton}, L.~W., {Kankelborg}, C.~C., {et~al.} 1999, \solphys,
  187, 229

\bibitem[{{Handy} {et~al.}(1998){Handy}, {Bruner}, {Tarbell}, {Title},
  {Wolfson}, {Laforge}, \& {Oliver}}]{1998SoPh..183...29H}
{Handy}, B.~N., {Bruner}, M.~E., {Tarbell}, T.~D., {et~al.} 1998, \solphys,
  183, 29

\bibitem[{{Howard} {et~al.}(1990){Howard}, {Harvey}, \&
  {Forgach}}]{1990SoPh..130..295H}
{Howard}, R.~F., {Harvey}, J.~W., \& {Forgach}, S. 1990, \solphys, 130, 295

\bibitem[{{Krijger} {et~al.}(2001){Krijger}, {Rutten}, {Lites}, {Straus},
  {Shine}, \& {Tarbell}}]{2001A&A...379.1052K}
{Krijger}, J.~M., {Rutten}, R.~J., {Lites}, B.~W., {et~al.} 2001, \aap, 379,
  1052

\bibitem[{{Leahy} {et~al.}(1983){Leahy}, {Darbro}, {Elsner}, {Weisskopf},
  {Kahn}, {Sutherland}, \& {Grindlay}}]{1983ApJ...266..160L}
{Leahy}, D.~A., {Darbro}, W., {Elsner}, R.~F., {et~al.} 1983, \apj, 266, 160

\bibitem[{{Linnell Nemec} \& {Nemec}(1985)}]{1985AJ.....90.2317L}
{Linnell Nemec}, A.~F. \& {Nemec}, J.~M. 1985, \aj, 90, 2317

\bibitem[{{Lites} \& {Chipman}(1979)}]{1979ApJ...231..570L}
{Lites}, B.~W. \& {Chipman}, E.~G. 1979, \apj, 231, 570

\bibitem[{{Lites} {et~al.}(1993){Lites}, {Rutten}, \&
  {Kalkofen}}]{1993ApJ...414..345L}
{Lites}, B.~W., {Rutten}, R.~J., \& {Kalkofen}, W. 1993, \apj, 414, 345

\bibitem[{{Muglach}(2003)}]{2003A&A...401..685M}
{Muglach}, K. 2003, \aap, 401, 685

\bibitem[{{O'Shea} {et~al.}(2001){O'Shea}, {Banerjee}, {Doyle}, {Fleck}, \&
  {Murtagh}}]{2001A&A...368.1095O}
{O'Shea}, E., {Banerjee}, D., {Doyle}, J.~G., {Fleck}, B., \& {Murtagh}, F.
  2001, \aap, 368, 1095

\bibitem[{Rutten(1995)}]{Rutten1995b}
Rutten, R.~J. 1995, in Helioseismology, ed. J.~T. Hoeksema, V.~Domingo,
  B.~Fleck, \& B.~Battrick, Proc.\ Fourth SOHO Workshop (ESA Publ.\ Div.,
  ESTEC, Noordwijk: ESA SP--376 Vol.\ 1), 151--163

\bibitem[{{Rutten} {et~al.}(2004){Rutten}, {de Wijn}, \& {S{\"
  u}tterlin}}]{2004A&A...416..333R}
{Rutten}, R.~J., {de Wijn}, A.~G., \& {S{\" u}tterlin}, P. 2004, \aap, 416, 333

\bibitem[{{Rutten} \& {Krijger}(2003)}]{2003A&A...407..735R}
{Rutten}, R.~J. \& {Krijger}, J.~M. 2003, \aap, 407, 735

\bibitem[{{Schmieder} \& {Mein}(1980)}]{1980A&A....84...99S}
{Schmieder}, B. \& {Mein}, N. 1980, \aap, 84, 99

\bibitem[{{Schwarzschild}(1948)}]{1948ApJ...107....1S}
{Schwarzschild}, M. 1948, \apj, 107, 1

\bibitem[{{Wunnenberg} {et~al.}(2002){Wunnenberg}, {Kneer}, \&
  {Hirzberger}}]{2002A&A...395L..51W}
{Wunnenberg}, M., {Kneer}, F., \& {Hirzberger}, J. 2002, \aap, 395, L51

\end{thebibliography}
\end{document}